\documentclass[twocolumn,english,showpacs,floatfix,amsmath,amssymb,prb,aps,superscriptaddress]{revtex4}
\usepackage[T1]{fontenc}
\usepackage[latin9]{inputenc}
\usepackage{babel}

\usepackage[caption=false]{subfig}
\usepackage{graphicx}
\usepackage{amssymb}
\usepackage{amsmath}

\newcommand{\be}{\begin{equation}}
\newcommand{\ee}{\end{equation}}
\newcommand{\ba}{\begin{eqnarray}}
\newcommand{\ea}{\end{eqnarray}}

\newcommand{\Gmixed}{\tilde{G}}
\DeclareMathOperator{\sgn}{sgn}

\begin{document}

\title{Heat transport through quantum Hall edge states: Tunneling
  versus capacitive coupling to reservoirs.}  \author{ Hugo Aita}

\affiliation{Departamento de F\'{\i}sica and IFLP, Facultad de
  Ciencias Exactas, Universidad Nacional de La Plata, cc 67, 1900, La
  Plata, Argentina.}

\author{ Liliana Arrachea}

\affiliation{Departamento de F\'{\i}sica and IFIBA, Facultad de
  Ciencias Exactas y Naturales, Universidad de Buenos Aires,
  Pabell\'on 1, Ciudad Universitaria, 1428, Buenos Aires, Argentina.}

\author{Carlos Na\'on}

\affiliation{Departamento de F\'{\i}sica and IFLP, Facultad de
  Ciencias Exactas, Universidad Nacional de La Plata, cc 67, 1900, La
  Plata, Argentina.}

\author{ Eduardo Fradkin}

\affiliation{Department of Physics and Institute for Condensed Matter
  Theory, University of Illinois at Urbana-Champaign, 1110 West Green
  Street, Urbana, Illinois 61801-3080, USA.}

\begin{abstract}
  We study the heat transport along an edge state of a two-dimensional
  electron gas in the quantum Hall regime, in contact to two
  reservoirs at different temperatures. We consider two exactly
  solvable models for the edge state coupled to the reservoirs. The
  first one corresponds to filling $\nu=1$ and tunneling coupling to
  the reservoirs. The second one corresponds to integer or fractional
  filling of the sequence $\nu=1/m$ (with $m$ odd), and capacitive
  coupling to the reservoirs. In both cases we solve the problem by
  means of non-equilibrium Green function formalism. We show that heat
  propagates chirally along the edge in the two setups. We identify
  two temperature regimes, defined by $\Delta$, the mean level spacing
  of the edge. At low temperatures, $T< \Delta$, finite size effects
  play an important role in heat transport, for both types of
  contacts.  The nature of the contacts manifest themselves in
  different power laws for the thermal conductance as a function of
  the temperature.  For capacitive couplings a highly non-universal
  behavior takes place, through a prefactor that depends on the length
  of the edge as well as on the coupling strengths and the filling
  fraction. For larger temperatures, $T>\Delta$, finite-size effects
  become irrelevant, but the heat transport strongly depends on the
  strength of the edge-reservoir interactions, in both cases. The
  thermal conductance for tunneling coupling grows linearly with $T$,
  whereas for the capacitive case it saturates to a value that depends
  on the coupling strengths and the filling factors of the edge and
  the contacts.

\end{abstract}

\date{\today}

\maketitle

\section{Introduction}
One of the most remarkable properties of the quantum Hall effect (QHE)
is the existence of topologically protected chiral edge
states. \cite{QH-edges} Originally unveiled by Laughlin \cite{laug}
and Halperin, \cite{Halperin-1982} the remarkable stability of these
states is a consequence of the peculiar chiral\cite{Halperin-1982,but}
and topological\cite{wen-niu} nature of the quantum Hall effect. After
the works of Wen, \cite{Wen-1990,Wen-1990b} and Kane and Fisher,
\cite{kane-fish} these states are viewed as realizations of a chiral
Luttinger liquid which is amenable to be investigated by means of
transport experiments.

The structure of the edge states reveals fundamental properties of the
quantum Hall state. At filling fraction $\nu=1$ it consists of a
single state located at the edge of the sample where electrons
propagate chirally.  Fractional quantum Hall states generally have a
more complex structure of edge states, with one or more edge
states. In general each edge state has a chirality (which can be
different from the other edge states) and its excitations carry
non-trivial quantum numbers such as (generally fractional) charge as
well as spin.  Some edge states do not carry charge (or spin)
excitations at all and are thus neutral.  Thus, edge states of
fractional quantum Hall states contribute in non-trivial ways to the
{\em charge} (and possibly {\em spin}) transport in the system.  In
addition, the edge states carry energy and hence contribute to the
thermal transport. These interesting features have been recently
investigated in systems in the integer and fractional Hall by means of
different thermometry techniques.  \cite{granger, bid, altimiras,
  yacoby,altimiras2,heiblum} In this paper we will focus on the {\em
  energy} (heat) transport properties of the edge states of the
simplest fractional quantum Hall states, the Laughlin states.

Experimental evidence of the chiral propagation of the heat along an edge state in
a GaAs/AlAs heterostructure with a two-dimensional electron gas in the
integer quantum Hall regime has been presented in
Ref. \onlinecite{granger}. The experiment was performed in the quantum
Hall regime with filling $\nu=1$ locally heated by injecting an ac
current from a source reservoir. These experimental features can be
captured by a simple one-dimensional model of non-interacting chiral
fermions connected to reservoirs through tunneling
couplings. \cite{us} It can be argued within that model that an
analogous chiral propagation of the heat is expected if a stationary
temperature gradient is applied between source and drain reservoirs,
instead of heating with an ac current.  Recent improvements in the
technology of the edge state manipulations also enable the possibility
of capacitive couplings. \cite{capcoup} These results show that tunneling and capacitive couplings can be controlled, separately, if the sizes of contacts are selected appropriately.  According to measurements in a quantum  Hall Fabry-Perot interferometer,\cite{mcclure} for a $18 \mu m^2$ device, Coulomb effects are not significant. The same study shows zero-bias oscillations in a $2 \mu m^2$ device of similar design, indicating (as expected) an increasing importance of charging effects in smaller samples.

While for the case of
tunneling coupling the heat current is accompanied with a particle
current, for capacitive couplings, the energy currents are isolated
from the particle flow.  This feature is interesting since it opens
the possibility for the study of energy and charge propagation
separately. Although in any realistic setup the tunneling coupling is always present, and it is always more relevant than the capacitive coupling, the experiments of Ref.[\onlinecite{capcoup}] show that it is possible to have a wide enough range of temperatures and voltages in which the tunneling coupling can be made small enough to be neglected.

Charge transport by tunneling coupling into the edge states has been
the subject of many theoretical works. A limited list of papers on the
topic is given by Refs.
[\onlinecite{QH-edges,kane-fish,cham-wen,cham-frad}].  Heat transport
along edge states has been considered in a smaller number of
studies.\cite{cappelli,grosfeld,kovrizhin} In addition to the work by
Kane and Fisher,\cite{kane-fis-97} we can mention
Ref.[\onlinecite{us}] which focuses on an ac driven edge corresponding
to a filling $\nu=1$ and follows the experimental work by Granger {\rm et
al.} \cite{granger} Another important recent work is
Ref.[\onlinecite{stern}], which is devoted to analyze thermoelectric
effects between edge states through a coupled quantum dot in a quantum
Hall bar with fillings $\nu=5/2$ and $\nu=2/3$.

The aim of the present work is to analyze heat transport induced by a
temperature gradient applied at reservoirs that are {\em capacitively
  coupled} to an edge state of a quantum Hall state with filling
$\nu$. We will consider the cases of an integer quantum Hall state,
with $\nu=1$, and of general Laughlin fractional quantum Hall states,
with filling fraction $\nu=1/m$, with $m$ odd. We solve this
problem exactly. For the particular case of filling $\nu=1$, we
compare with the behavior of the heat transport induced by a
temperature gradient at reservoirs connected to the edge by {\em
  tunneling} at point contacts, which is also an exactly solvable
problem. The more general case which involves tunneling at point
contacts is not exactly solvable and will be discussed elsewhere.  In
all the cases we focus on two properties: a) the thermal conductance
of the edge, and b) the behavior of the local temperature along the
non-equilibrium edge. The latter is defined by recourse to a
``thermometer'', which is realized by a third reservoir which is very
weakly coupled to the edge. The temperature of this reservoir is such
that the heat flow through the contact vanishes.  In this paper we
show that there is a different qualitative behavior of the heat
conductance for tunneling and capacitive couplings.  The behavior of
the local temperature is, however, very similar in both cases.  The
local temperature displays a profile with discontinuities at the
contacts to the reservoirs, indicating that the edge tends to
thermalize with the closest upstream reservoir.

The paper is organized as follows. In Section II we introduce the two
models to be exactly solved and define the heat currents in both
cases. In Section III we discuss the energy balance along the devices
and give explicit formal expressions for heat currents in terms of
correlators. In Section IV we present the calculations of the heat
transport using a Keldysh non-equilibrium Green function formalism. In
Section V we present results for the behavior of the local temperature
along the edge as well as the thermal conductance. Section VI is
devoted to summary and conclusions. Finally, in Appendices A and B we
gather some details of the calculations.

\begin{figure}
  \centering
  \includegraphics{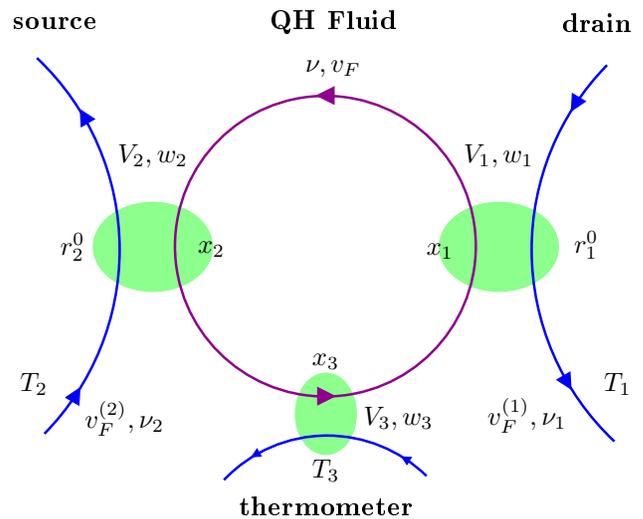}
  \caption{(Color online) Sketch of the studied setup of a fractional
    (Laughlin) quantum Hall fluid in contact with a source and a drain
    and with a thermometer.  The fractional quantum Hall edge state is
    represented by a ring. Two reservoirs, drain and source, with
    temperatures $T_1 < T_2$, are connected to the ring at positions
    $x_1$ and $x_2$, either through contacts that allow tunneling of
    particles with couplings strengths $w_1$, $w_2$, or through
    capacitive couplings with strengths $V_1$, $V_2$ where only an
    energy current can flow through the leads. A third reservoir is
    weakly connected at $x_3$ in order to sense the local temperature
    $T_3$. For tunneling contacts, the only exactly solvable case
    corresponds to filling $\nu=1$. For capacitive coupling, any
    filling $\nu$ can be exactly treated.}
  \label{fig1}
\end{figure}

\section{Models}

The full system under consideration is sketched in Fig.\ref{fig1}. It
is described by the following Hamiltonian \be
\label{ham} H=H_{edge}+\sum_{\alpha=1}^M [H_{\alpha}+H_{c,\alpha}].
\ee The edge states of a quantum Hall fluid are represented by a
one-dimensional (1D) system, a ring of circumference $L$, along which
chiral fermions circulate with velocity $v_F$. The edge of the fluid
is coupled to $M$ reservoirs (the leads) of infinite length, which are being represented
by a set of ``edge states'' with fillings $\nu=1$, for tunneling coupling, or belonging to the Laughlin series, in the capacitive case. We will focus, in
particular, on the configuration sketched in Fig. \ref{fig1}, with
$M=3$ reservoirs. The drain ($\alpha=1$) and source ($\alpha=2$)
reservoirs are at temperatures $T_1$ and $T_2$, respectively, with
$T_2>T_1$. The third reservoir corresponds to a thermometer which
senses the local temperature $T_3$. The latter is defined from the
condition of local thermal equilibrium, implying a vanishing heat flow
between this reservoir and the edge.  We work in units where
$\hbar=k_B=e=1$, but we will restore this universal constants in the
discussion of the results.

\subsection{Tunneling coupling}
\label{sec:model}

In this case we focus on a quantum Hall state with integer filling
$\nu=1$, which is the only exactly solvable case for the present
geometry and under the effects of a temperature gradient.  The
Hamiltonian is \ba
H_{edge} &= &- i v_{F } \int_{0}^L dx : \Psi^{\dagger}(x) \partial_x \Psi(x) : \nonumber \\
& = & \sum_p v_{F}\, p\, c^{\dagger}_p c_p, \ea where $p= 2 n
\pi/L$, with $n$ an integer.  A large upper (UV) momentum cutoff
$\Lambda=D/v$ (where $D$ is the bandwidth of the edge states) will be
assumed.  The bandwidth $D$ will also be assumed to be small compared
to the gap that separates the Landau levels.

We represent the reservoirs by systems of 1D chiral fermions of length
$L_{\alpha}$, which we will assume to be thermodynamically large,
$L_{\alpha} \rightarrow \infty$.  The corresponding Hamiltonian
$H_{\alpha}$ for each of these systems reads \be H_{\alpha}=-i
v_{F}^{\alpha} \int_0^{L_{\alpha}} d r_{\alpha}
\Psi^{\dagger}(r_{\alpha}) \partial_{r_{\alpha}} \Psi(r_{\alpha}).
\ee The source and drain reservoirs as well as the thermometer have
the same chemical potential $\mu$.

The contacts are described by the Hamiltonians \be \label{cont}
H_{c,\alpha} = w_{\alpha} \sum_{\sigma}
[\Psi^{\dagger}_{\sigma}(x_{\alpha}) \Psi_{\sigma}(r^0_{\alpha})+ H.
c.], \ee where $x_{\alpha}$ and $r^0_{\alpha}$ are, respectively, the
positions of the ring and the reservoir at which the contact is
established. We assume that the tunneling parameter $w_3$ between the
ring and the thermometer is so weak that it introduces negligible
dephasing in the particle propagation along the ring.

\subsection{Capacitive coupling}

In this section we define the model corresponding to a capacitive
coupling between the edge and the reservoirs. Assuming a local
coupling, this corresponds to considering the Hamiltonian of
Eq.\eqref{ham} with the terms $H_{c,\alpha} = V_{\alpha}
\Psi^{\dagger} (x_{\alpha}) \Psi(x_{\alpha})
\Psi^{\dagger}(r^0_{\alpha}) \Psi(r^0_{\alpha})$, where $r^0_{\alpha}$
and $x_{\alpha}$ are coordinates of the reservoir and the edge,
respectively.  As it is well-known, this type of quartic interactions
can be more easily handled by adopting a bosonic representation of
the edge states.\cite{Stone} In that language, the fermionic density
becomes proportional to the spatial derivative of a free chiral
bosonic field $\phi(x)$, which represents a quantum fluctuation
propagating along the edge of the quantum Hall fluid. Since the bulk
of the quantum Hall fluid is gapped (and hence incompressible), the
edge of the fluid can be regarded as a ring of finite length $L$ of
non-interacting chiral bosons (with a fixed ``compactification
radius'' determined by the filling fraction of the bulk quantum Hall
fluid, see Ref.[\onlinecite{QH-edges}]) capacitively coupled to
reservoirs at different temperatures. The reservoirs are also
described by 1D chiral bosons of infinite length, with fillings $\nu_{\alpha}$.

The total Hamiltonian has the structure of Eq. \eqref{ham}. In the
bosonized language the Hamiltonian for the edge is given by
\begin{equation}
  \label{hbose}
  H_{edge} = \frac{v_{F}}{4 \pi \nu}
  \int_0^L dx :(\partial_x \phi(x) )^2: + \frac{\pi}{L}
  \hat{N}(\hat{N}+1),
\end{equation}
where $\hat{N}$ is the number operator corresponding to the original
fermionic system (See Ref. [\onlinecite{Haldane}] for details) and
$\nu$ is the filling fraction. The present case can be solved for a
quantum Hall state with a filling fraction $\nu$, which can be integer
as well as fractional with the law $\nu=1/m$, with $m$ odd.

The Hamiltonians for the reservoirs read
\begin{eqnarray}
  H_{\alpha} & = &  \frac{v_{F}^{\alpha}}{4 \pi \nu_\alpha} \int_0^{L_{\alpha}}
  dr_{\alpha}  : (\partial_{r_{\alpha}} \phi(r_{\alpha}) )^2 :.
\end{eqnarray}
As in the tunneling case, we will consider the leads to be infinitely
long, $L_{\alpha} \rightarrow \infty$.  The contact between the
central system and the two reservoirs is
\begin{equation}
  H_{c,\alpha}= V_{\alpha} \partial_{r_{\alpha}} \phi(r_{\alpha}) |_{r_{\alpha}= r_{\alpha}^0} \partial_{x} \phi(x) |_{x= x_{\alpha}},
\end{equation}
where $x_{\alpha}$ and $r^0_{\alpha}$ are the points on the ring and
the reservoir, respectively, that intervene in the coupling.

The chiral Bose fields $\phi(x)$ and $\phi(r_{\alpha})$ satisfy the
equal-time commutation relations
\begin{eqnarray}
  \label{boscom}
  \left[\phi(r_{\alpha}),\phi(x) \right] &=& 0 \\
  \left[\phi(x), \phi(x^{\prime}) \right] &=& - i \pi \nu \sgn(x-x^{\prime}) \\
  \left[\phi(r_{\alpha}), \phi(r^{\prime}_{\alpha}) \right] &=&
  - i\pi \nu_\alpha \sgn(r_{\alpha}-r^{\prime}_{\alpha}).
\end{eqnarray}

\section{Energy balance and heat current}

Our aim is to evaluate the heat current flowing through the contacts
between the edge state and a given reservoir $\alpha$.  To this end we
analyze the time dependence of the energy stored in the reservoir.  In
the case of the tunneling coupling we consider \ba \dot{Q_{\alpha}}&=&
\dot{E_{\alpha}}-\mu \dot{N_{\alpha}}= -i \langle [H_{\alpha}-\mu
N_{\alpha}, H_{c,\alpha}] \rangle = J^{Q, {\rm t}}_{\alpha}\nonumber\\
&& \ea where $E_{\alpha}$ and $N_{\alpha}$ are, respectively the
energy and the charge stored in the reservoir $\alpha$.  In order to
relate energy flow to heat flow we subtracted the convective component
$\mu \dot{N_{\alpha}}$.  The result is \be
\label{curt0}
J^{Q, {\rm t}}_{\alpha} = - 2 \mbox{Re} \Big\{ \int \frac{d
  p_{\alpha}}{2 \pi} w_{p_{\alpha}} (\varepsilon_{p_{\alpha}} - \mu)
{\tilde G}^<(x_{\alpha},p_{\alpha};t,t) \Big\}, \ee where
$\varepsilon_{p_{\alpha}}= v_F^{\alpha} p_{\alpha}$, and
$w_{p_{\alpha}} = w_{\alpha} e^{-i p_{\alpha} r_{\alpha}^0}
/\sqrt{L_{\alpha}}$. The lesser Green function is \be
\label{lesg}
{\tilde G}^<(x_{\alpha},p_{\alpha};t,t^{\prime})= i \langle
c^{\dagger}_{p_{\alpha}}(t^{\prime}) \Psi(x_{\alpha},t) \rangle.  \ee

In the case of the capacitive coupling there is no particle
flow. Thus, the energy flow is equivalent to the heat flow
\begin{equation}
  \dot{Q_{\alpha}}=\dot{E_{\alpha}}= -i \langle [H_{\alpha}, H_{c,\alpha}] \rangle = J^{Q,{\rm c}}_{\alpha}.
\end{equation}
The calculation yields
\begin{eqnarray}
  \label{hcur}
  J^{Q, {\rm c}}_{\alpha} & = & i V_{\alpha} v_F^{\alpha}
  \partial_x  \partial^2_{r_{\alpha}} {\tilde D}^<(x,r_{\alpha};t,t)|_{x=x_{\alpha}, r=r_{\alpha}^0},
\end{eqnarray}
with the lesser function defined as
\begin{equation}
  \label{lesd}
  {\tilde D}^<(x,r_{\alpha};t,t^{\prime})= i
  \langle \phi(r_{\alpha})(t^{\prime}) \phi(x,t) \rangle.
\end{equation}
Notice that $[\hat{N}, \partial_x \phi(x)]=0$, thus the last term of
(\ref{hbose}) does not contribute to the heat current.

\section{Methodology: non-equilibrium Green functions}
\subsection{Tunneling coupling}
In order to compute the current we must evaluate the lesser Green
function given in Eq.\eqref{lesg}. To this end, we define the retarded
Green function \ba
\label{gxx}
& & G^R(x,x^{\prime};t,t^{\prime}) = - i \Theta(t-t^{\prime}) \langle
\{ \Psi(x,t) ,
\Psi^{\dagger}(x^{\prime},t^{\prime}) \} \rangle \nonumber\\
&& \ea where $x,x^{\prime}$ are coordinates on the ring. This is a
rather standard procedure which we summarize for completeness in the
Appendix \ref{apa}. The lesser Green function entering the expression
of the current of Eq.\eqref{curt0} can be calculated from
Eqs. \eqref{dyles-tun}. The result is
\begin{widetext}
  \ba J^{Q, {\rm t}}_{\alpha}& = & - \int_{-\infty}^{+\infty} \frac{d
    \omega}{2 \pi} (\omega - \mu) \Gamma^{\rm t}_{\alpha}(\omega)
  \left[ 2 \mbox{Im}[G^R(x_{\alpha},x_{\alpha}; \omega)
    f_{\alpha}(\omega) ] + \sum_{\beta=1}^M |G^R(x_{\alpha},x_{\beta};
    \omega)|^2 \Gamma^{\rm t}_\beta(\omega) f_{\beta}(\omega)
  \right], \label{curtt} \ea
\end{widetext}
where the function $G^R(x_{\alpha},x_{\beta}; \omega)$ is obtained
from the second equation of the set of Eq.\eqref{dyret-tun} and
$f_{\alpha}(\omega)$ is the Fermi-Dirac function which depends on the
chemical potential and temperature of the reservoir $\alpha$. The
hybridization function $\Gamma^{\rm t}_{\alpha}(\omega)$, defined in
Eq. \eqref{gammat}, depends on the density of states of the reservoir,
which in our case is a constant within the bandwidth characterized by
an energy cutoff $\Lambda$, and the square of the tunneling amplitude
$|w_{\alpha}|^2$ between the edge and the reservoir.

An alternative representation for this current is obtained by
substituting the identity of Eq.\eqref{idt} into Eq.\eqref{curtt}.
The resulting expression reads \be
\label{jqtun}
J^{Q,{\rm t}}_{\alpha} = \sum_{\beta=1}^M \int_{-\infty}^{+\infty}
\frac{d \omega }{2 \pi} (\omega - \mu){\cal T}^{\rm t}_{\alpha,
  \beta}(\omega)[f_{\alpha}(\omega)-f_{\beta}(\omega)], \ee which has
the familiar form of a Landauer-B\"uttiker formula.  The heat current
resulting from a difference of temperatures imposed at the reservoirs
is expressed in terms of the corresponding difference of Fermi
functions times the amount of heat transferred by the tunneling of
particles, $\omega - \mu$, times the transmission function which
quantifies the transparency of the system in contact to the
reservoirs. The latter function in our case reads
\begin{equation} {\cal T}^{\rm t}_{\alpha, \beta}(\omega)= \Gamma^{\rm
    t}_\alpha(\omega)|G^R(x_{\alpha},x_{\beta}; \omega)|^2 \Gamma^{\rm
    t}_\beta(\omega)
\end{equation}
which depends on the Green function of the coupled edge and the
hybridization functions of the coupled reservoirs.

A typical plot for the transmission function ${\cal T}^{\rm
  t}_{1,2}(\omega)$ of a two terminal setup is shown in
Fig. \ref{fig3} .  It is evaluated by solving the set of two coupled
equations defined by Eq. \eqref{dyret-tun} for $M=2$ reservoirs and
$x^{\prime}=x_{\alpha}$, with $\alpha=1,2$.  The result is a sequence
of resonances which define Lorentzian peaks at the positions
$\varepsilon_{k_n}= v_F 2 \pi n/L$ of the energies of the isolated
edge. The hybridization to the reservoirs generate finite lifetime of
the electrons occupying those states, which is accounted by the width
$\propto \Gamma^{\rm t}_{\alpha}$ of the peaks of the transmission
function. For the present model of reservoirs, the widths as well as
the heights of the resonant peaks, are constant.

\subsection{Capacitive coupling}

In this case the evaluation of the heat current of Eq.\eqref{hcur}
requires the computation of the lesser function which is given by
Eq.\eqref{lesd}. We define the retarded Green function
\begin{eqnarray}
  D^R(x,x^{\prime};t, t^{\prime}) & = &  -i \Theta(t-t^{\prime})
  \langle [ \phi(x,t) , \phi(x^{\prime}, t^{\prime}) ] \rangle,
\end{eqnarray}
In Appendix \ref{apb} we present the calculation of the corresponding
Dyson equations. Upon substituting Eq.\eqref{der2} into
Eq.\eqref{hcur}, we get the explicit expression
\begin{widetext}
  \begin{eqnarray}
    J^{Q, {\rm c}}_{\alpha} & = &-\frac{1}{2}
    \int_{-\infty}^{+\infty} \omega \Gamma^{\rm c}_{\alpha}(\omega)
    \left[
      2 \mbox{Im}[{\cal D}^R(x_{\alpha},x_{\alpha};\omega)]  n_{\alpha}(\omega) +
      \sum_{\beta=1}^M
      |{\cal D}^R(x_{\alpha},x_{\beta};\omega)|^2
      \Gamma^{\rm c}_{\beta}(\omega)  n_{\beta}(\omega)
    \right],
  \end{eqnarray}
\end{widetext}
where the retarded Green function ${\cal D}^R(x,x^{\prime};\omega)$ is
defined in Eq.\eqref{cald}.  Using the identity given in
Eq.\eqref{equal}, the heat current can be expressed as \be
\label{jqcap}
J^{Q, {\rm c}}_{\alpha} = \sum_{\beta =1}^M \int_{-\infty}^{+\infty}
\frac{d\omega}{2 \pi} \omega {\cal T}^{\rm
  c}_{\alpha,\beta}(\omega)[n_{\alpha}(\omega)-n_{\beta}(\omega)], \ee
where we have defined the transmission function
\begin{equation} {\cal T}^{\rm c}_{\alpha,\beta}(\omega)= \Gamma^{\rm
    c}_{\alpha}(\omega) |{\cal D}^R(x_{\alpha},x_{\beta};\omega) |^2
  \Gamma^{\rm c}_{\beta}(\omega)/2
\end{equation}
This function has the same properties as its tunneling counterpart,
${\cal T}^{\rm t}_{\alpha,\beta}(\omega)$. In particular, it satisfies
the symmetry
\begin{equation} {\cal T}^{\rm c}_{\alpha,\beta}(\omega)={\cal T}^{\rm
    c}_{\beta,\alpha}(\omega),\;\forall \;\alpha, \beta
\end{equation}
which implies the continuity of the heat current.

\begin{figure}[hbt]
  \centering \subfloat[][Tunneling coupling]
  { \includegraphics{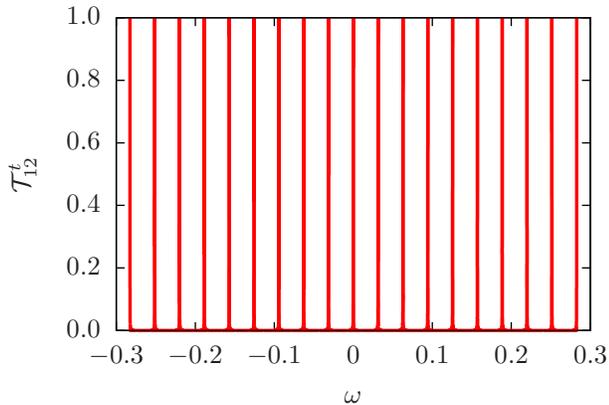} \label{fig3}}

  \subfloat[][Capacitive coupling]{
    \includegraphics{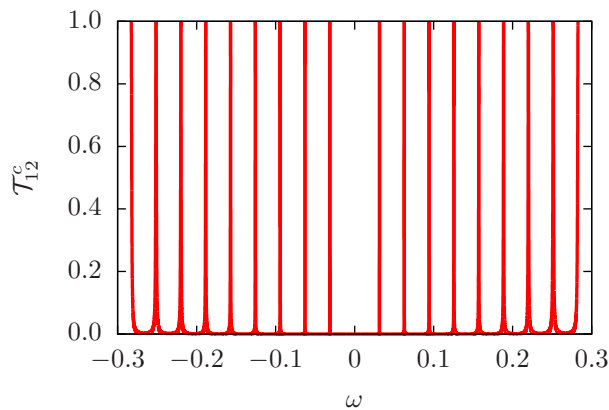}\label{fig4}}

  \caption{(Color online) Upper panel: Transmission function
    $\mathcal{T}_{12}{\rm t}$ as function of $\omega$ for a tunneling
    coupling between the edge and two reservoirs with amplitudes
    $w_{1} = w_{2} =0.1$.  Lower panel: Transmission function
    $\mathcal{T}_{12}{\rm c}$ as function of $\omega$ for a capacitive
    coupling between the edge and two reservoirs with amplitudes
    $V_{1} = V_{2} =0.2$.  The remaining parameters, common to the two
    cases, are $v_F=1$ and $L=200$, $x_1=0$ and $x_2= 100$. All
    energies are expressed in natural units ($\hslash=1$).}
\end{figure}

The above expression for the heat current has the same structure as
the corresponding one for tunneling contacts given in
Eq. \eqref{jqtun}. The temperature difference imposed at the
reservoirs enters in the present case in the Bose-Einstein functions
instead of in the Fermi-Dirac ones. In the present case, there is no
flow of particles. Thus, the energy transferred $\omega$ is directly
interpreted as heat. The transmission function depends on the
amplitude of the capacitive couplings as well as on the spectral
function of the bosonic reservoirs through the functions $\Gamma^{\rm
  c}_{\alpha}(\omega) $ defined in Eq. \eqref{gammac}.

A typical plot of the transmission function in a two-terminal
configuration is shown in Fig. \ref{fig4} . In the present case, we
must evaluate the linear set of two equations defined by
Eq. \eqref{dys4} with $M=2$ for $x^{\prime}=x_{\alpha}, \;
\alpha=1,2$.  As in the tunneling case, the result consists of a set
of resonances with a spacing $\Delta \omega \sim 2 \pi v_F/L$
corresponding to the energies of the uncoupled ring.  The coupling to
the reservoirs introduces a finite lifetime which determines the width
$\propto | \omega| V_{\alpha}^2 $ of the peaks of the transmission
function.  As in the case of the tunneling coupling, the height of
these peaks achieves the maximum value, equal to one, at resonance.
However, at low energies $|\omega|<2 \pi v_F /L$, there is a strong
suppression of the spectral weight. As we will discuss in the next
section, this effect renders the transmission of heat vanishing small
for low temperatures.

\section{Results}

\begin{figure}
  \centering \subfloat[][Tunneling coupling]
  {\includegraphics{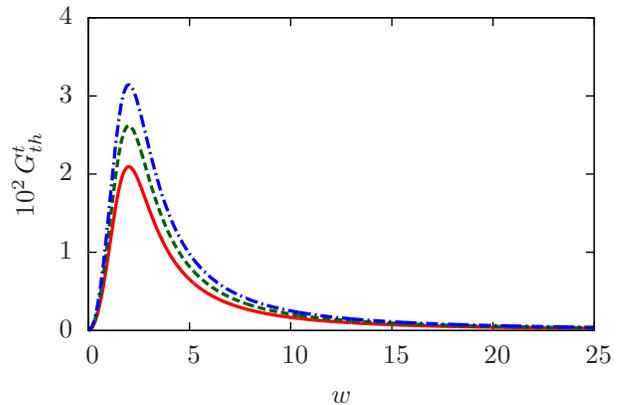} \label{fig9}}

  \subfloat[][Capacitive coupling]
  {\includegraphics{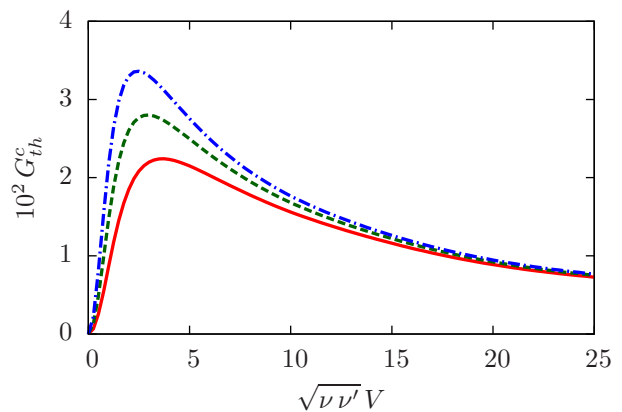}\label{fig10}}

  \caption{(Color online) Behavior of the thermal conductance as a
    function of the coupling strength within the regime $k_B T \gg
    \Delta$.  Upper panel: Tunneling thermal conductance $G^t_{th}$,
    having set $w_1 = w_2 = w$, for different temperatures: $T=0.04$
    (red, solid), $T=0.05$ (green, dashed), $T=0.06$ (blue,
    dot-dashed)).  Lower panel: Capacitive thermal conductance
    $G^c_{th}$ as function of $\sqrt{\nu \nu^\prime} V$ for different
    temperatures: $T=0.04$ (red, solid), $T=0.05$ (green, dashed),
    $T=0.06$ (blue, dot-dashed). We have set $V_1 = V_2 = V$ and
    $\nu_1 = \nu_2 = \nu^\prime$, and $\nu$ is the filling factor of
    the ring.  All energies are expressed in natural units
    ($\hslash=1.$).}
\end{figure}

We now turn to the discussion of the behavior of the thermal transport
through the edge state.  We will analyze the thermal conductance and
the local temperature sensed by a thermometer non-invasively coupled
along the edge. We consider a two terminal configuration with
reservoirs at different temperatures, $T_1$ and $T_2$ connected,
respectively, at $x=x_{1}$ and $x=x_{2}$.

\subsection{Thermal conductance}

We consider the source and drain reservoirs at slightly different
temperatures $T_1=T$ and $T_2= T+ \delta T$.  The thermal conductance
of the coupled edge state reads
\begin{equation}
  \label{eq:thermal-cond-def}
  G_{th} = \lim_{\delta T \to 0}
  \frac{J^Q (T+\delta T) -
    J^Q(T) }{\delta T},
\end{equation}
where $J^Q$ is the heat current flowing through the contacts between
the edge and the reservoirs. Notice that it is the same for the two
contacts because of the continuity of the energy and charge flows.  In
the forthcoming discussion presented within this section, it is
convenient to restore the units in terms of the fundamental constants
$h,\;e, \; k_B$.

\subsubsection{Tunneling coupling}

For tunneling coupling, we can recognize two different regimes: the
mesoscopic case ($k_B T \ll \Delta$) and the macroscopic one ($T \gg
\Delta$), where $\Delta= \hbar v_F 2 \pi/L $ is the level spacing of
the edge. In the former, the conductance depends not only on $T$ but
also on the chemical potential $\mu$. We distinguish two different
situations depending on wether $\mu$ coincides with one of the energy
levels of the edge (resonance) or it lies between two energy levels
(off-resonance). We hereafter focus on small $k_B T/ \mu$ where we can
resort to Sommerfeld expansion in Eq. \eqref{jqtun} provided that we
are in a regime where the transmission function is slowly varying with
$\omega$. This expansion casts \be J^{Q, {\rm t}}= \frac{k_B^2
  \pi^2}{3 h} \frac{d}{d\omega}\left[(\omega-\mu) {\cal T}^{\rm
    t}_{2,1}(\omega)\right]|_{\omega=\mu}\,T \,\delta T , \ee which
yields a linear behavior for the thermal conductance
\begin{equation}
  G_{th}^{\rm t}(T)= \kappa_0  {\cal T}^{\rm t}_{2,1}(\mu) T,
\end{equation}
where $\kappa_0= k_B^2 \pi^2/3 h$ is the universal thermal conductance
quantum constant.  As shown in Fig. \ref{fig3} , the transmission
function has a structure with several peaks and can be approximated by
the constant ${\cal T}^{\rm t}_{2,1}(\mu)$ only in two limits. The
first one corresponds to a resonant $\mu$ and the ultra small range of
temperatures $k_B T \ll \gamma $, being $\gamma$ the width of the
peaks, where the transmission function is ${\cal T}^{\rm
  t}_{2,1}(\omega) \sim 1$. The second one corresponds to an
off-resonant $\mu$ and also a small range $k_B T \ll \Delta$, where
the transmission function is ${\cal T}^{\rm t}_{2,1}(\omega) \sim 0$.

In the macroscopic regime, $k_B T \gg \Delta$, the conductance also
grows linearly with $T$ and does not depend on $\mu$,
\begin{equation}
  \label{gtun-highT}
  G_{th}^{\rm t} = f(\tilde{w}_{1},\tilde{w}_{2})\,\kappa_0 \,T,
\end{equation}
where the function $f(\tilde{w}_{1},\tilde{w}_{2})$ has the form
\begin{equation}
  \begin{split}
    \label{fw1w2}
    f(\tilde{w}_{1},\tilde{w}_{2}) &= \frac{4 \tilde{w}_{1}^2
      \tilde{w}_{2}^2} {\tilde{w}_{1}^2
      +\tilde{w}_{2}^{2}}\\
    & \quad \times \frac{1} {\left(1+\tilde{w}_{1}^{2}
        \tilde{w}_{2}^{2} \right)},
  \end{split}
\end{equation}
where we have defined $\tilde{w}_\alpha = w_\alpha/2\hslash \sqrt{v_F
  v_F^\alpha}$.  In Appendix \ref{app:analytic} we present an analytic
derivation of this result. Interestingly, this implies that, in this
regime, $ G_{th}^{\rm t}$ has a non-monotonic behavior as function of
the ring-reservoir coupling strength, as shown in Fig. \ref{fig9}.  A
similar behavior has been previously found in
Ref.\onlinecite{Karevski-Platini} for the magnetization current in a
XX spin-1/2 chain coupled to quantum reservoirs, and in steady
state thermal current in an open XY spin-1/2
chain.\cite{Prosen-Zuncovic} We have also verified that a similar
behavior takes place for the thermal conductance of a tight-binding
chain connected to one-dimensional electron reservoirs through a
tunneling coupling with a mismatching.  The fact that the conductance
at a fixed $T$ grows as a function of the coupling to the reservoirs
until a maximum value and then decreases for even larger couplings is
a priori non-intuitive. Interestingly, it is a consequence of the
coherent nature of the heat propagation. In fact, notice that the
quantity $v^{\alpha}_{\rm t} = w_{\alpha}/2\hbar$ can be interpreted
as the velocity with which the electrons travel through the tunneling
coupling, while the quantity $\tilde{w}_\alpha^2= (v^{\alpha}_{\rm
  t}/v_F^{\alpha})(v^{\alpha}_{\rm t}/v_F)$, entering in
Eq.\eqref{fw1w2} is a measure of the velocity mismatch for the
electron motion through the junction, the one within the reservoir and
the one along the ring.

The behavior of Fig.  \ref{fig9} shows that a small thermal flow
between the two reservoirs is expected for a high mismatching between
these three velocities.  This may occur for a very weak coupling
$w_{\alpha}$ in which case the velocity of tunneling is much smaller
than the velocities that the electrons have within the reservoirs and
within the finite-size central edge. A similar effect is expected for
a large $w_{\alpha}$, which corresponds to $v^{\alpha}_{\rm t} \gg
v_F^{\alpha}, \; v^{\alpha}_{\rm t} \gg v_F$. In this case the
electrons jump through the contact at a much higher velocity than the
one with which they propagate within the reservoirs and along the
central ring, resulting in a poor net transmission from one reservoir
to the other. It is important to notice that the function
$f(\tilde{w}_{1},\tilde{w}_{2}) \le 1$. The thermal conductance is
thus upper bounded by its ballistic value $\kappa_0 T$ and satisfies
the limit set in Ref. \onlinecite{pendry}.  An alternative heuristic
derivation of this limit consists in requesting that the thermal
conductance satisfies the uncertainty principle $\Delta E \tau \geq
\hbar/2 $, where $\Delta E \sim k_B \delta T$ and $\tau= k_B
T/J^{Q,{\rm t}}$. Using $J^{Q,{\rm t}}= G_{th}^{\rm t} \delta T$, we
get $G_{th}^{\rm t} \leq 2 k_B^2 T /\hbar$, which is approximately the
exact upper bound.

\begin{figure}
  \centering
  \includegraphics{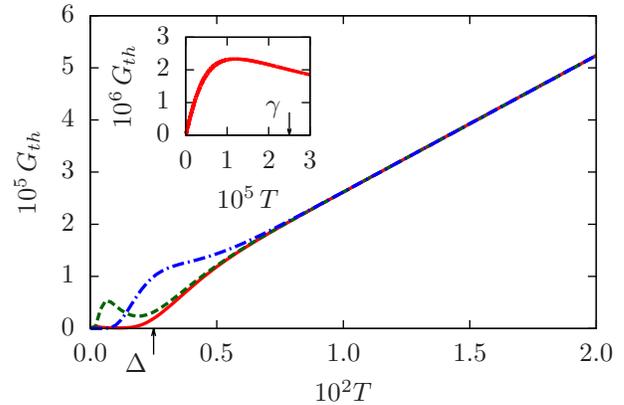}
  \caption{(Color online) Low temperature behavior of the tunneling
    thermal conductance through reservoirs $1$ and $2$ as function of
    temperature, for different values of the chemical potential of the
    reservoirs: $\mu = 0$ corresponding to resonance (red, solid), and
    two off-resonant values $\mu=\pi/(5L)$ (green, dashed), $\pi/L$
    (blue, dot-dashed)).  The couplings are $w_{1}=w_{2}=0.1 $, and
    the ring length $L=400$. The arrows indicate $T=\gamma, \;
    \Delta$.  A zoom of the linear regime for very low temperatures
    ($T \sim \gamma$) in the resonant case is shown in the inset. }
  \label{fig5}

\end{figure}

\begin{figure}
  \centering
  \includegraphics{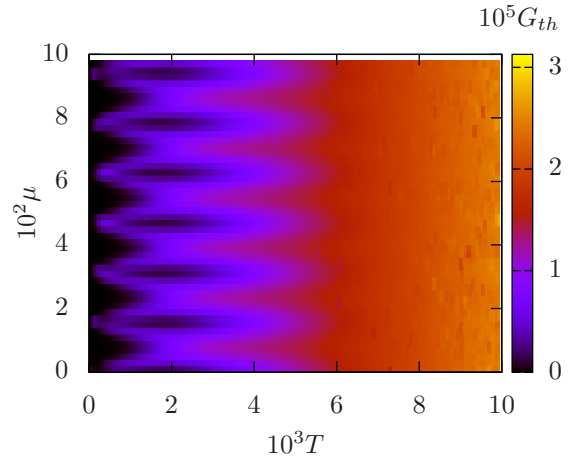}
  \caption{(Color online) Map plot of the tunneling thermal
    conductance through reservoirs $1$ and $2$, as function of the
    temperature and the chemical potential of the reservoirs.  The
    couplings are $w_{1}=w_{2}=0.1 $, and the ring length $L=400$.}
  \label{mapplot}
\end{figure}

The different regimes are illustrated in Fig. \ref{fig5}. The solid
red plot corresponds to a resonant $\mu$ and we can distinguish a very
narrow region close to $T=0$ where $G_{th}^{\rm t}(T)$ grows linearly
(see the inset of the figure). For $T>\gamma$ there is a structure
related to the end of the peak, while for $T>\Delta$ it grows again
linearly. In the other plots, corresponding to off-resonant $\mu$, it
can be seen a vanishing conductance for $T < |\mu-\varepsilon_c|$
where $\varepsilon_c$ is the energy of the energy level of the edge
that is the closest one to $\mu$. A structure (maximum) related to the
existence of a neighboring peak is observed and for larger $T$ the
conductance is again linear. This behavior is repeated as $\mu$ is
varied through the sequence of peaks and valleys, as shown in the
contour plot of Fig. \ref{mapplot}.

To summarize, the universal behavior of the thermal
conductance\cite{kane-fis-97} \be G^{\rm c}_{th}= \kappa_0 T \ee is
expected in the limit of ultra low temperature $k_B T \ll \gamma$ and
for a resonant $\mu$. In the high temperature regime $k_B T \gg
\Delta$ the conductance becomes independent of $\mu$ and grows again
linearly, but the slope is not universal and depends on the coupling
constants as described by Eqs. \eqref{gtun-highT} and \eqref{fw1w2}.

\subsubsection{Capacitive coupling}

In this case, the heat current is given by Eq. \eqref{jqcap}, which
depends on a difference of Bose-Einstein distribution functions. The
detailed behavior depends on the spectral properties described by the
transmission function ${\cal T}^{\rm c}_{12}(\omega)$ at low frequency
$\omega$. As shown in the lower panel of Fig. \ref{fig3}, this
function consists in a set of Lorentzian peaks centered at the
energies of the uncoupled ring. Thus, for low enough temperatures,
smaller than the mean level spacing $\Delta= 2 \pi \hbar v_F/ L$,
($k_B T < \Delta $) we can perform an expansion of the transmission
function around $\omega=0$.  It results in the frequency dependence
\begin{equation} {\cal T}^{\rm c}_{12}(\omega) \sim \gamma_c \omega^2
\end{equation}
The low temperature behavior for the heat current is then described by

\be J^{Q,{\rm c}} = \gamma_c \int_{-\infty}^{+\infty} d \omega
\omega^3 \left[ n(T_1)-n(T_2) \right].  \ee The above integral can be
evaluated in the limit of $T \rightarrow 0$ following standard
procedures,\cite{pathria} leading to the result
\begin{equation}
  J^{Q,{\rm c}} =( \gamma_c k_B^4 \pi^2/30) \left( T_1^4-
    T_2^4 \right)
\end{equation}
which implies the following law for the low-temperature behavior of
the thermal conductance
\begin{equation}\label{G-cubic}
  G_{th}^{\rm c}(T) =\kappa_1 T^3, \;\;\;\;\;\;\;\;\;\;\;\;\;k_B T\ll \Delta,
\end{equation}
with
\begin{eqnarray}
  \kappa_1 &=&  \kappa_0 k_B^2  \lambda_1, \nonumber \\
  \lambda_1 & = &  \frac{128 \pi^6 }{5 }
  \frac{ \hslash^2 v_F^2}{L^2}
  ( \tilde{V}_1 \tilde{V}_2 )^2,
\end{eqnarray}
and $\tilde{V}_{\alpha} = \sqrt{\nu \nu_\alpha} V_{\alpha}/(\hslash
v_F) (\hbar v_F^{\alpha})$.  For higher temperatures ($k_B T \gg
\Delta$), the analysis is more subtle. An analytical computation can
be performed by considering an approximate form of (\ref{jqcap}),
valid in this macroscopic regime. We describe this approach in
Appendix \ref{app:analytic-capacitive}. In contrast to the tunneling
case, here an intermediate regime may emerge provided that the
coupling strengths $V_{1}$ and $V_{2}$ satisfy $1/\sqrt{\tilde{V}_{1}
  \tilde{V}_{2}} \gg \Delta$.  Under this condition we can distinguish
a regime where the thermal conductance follows again a cubic power
law, but with a prefactor that is independent of the length of the
edge,
\begin{equation}
  G_{th}^{\rm c}(T) = \kappa_2 \, T^3, \;\;\;\;\;\;\;\;\;\;\;\;\; \Delta \ll k_B T \ll 1/\sqrt{\tilde{V}_{1} \tilde{V}_{2}},
\end{equation}
with
\begin{eqnarray}
  \label{kappa-cap2}
  \kappa _2& =&  \kappa_0 k_B^2  \lambda_2, \nonumber \\
  \lambda_2 & = &  \frac{32 \pi^4 }
  {5}\, \frac{\tilde{V}_1^2 \, \tilde{V}_2^2}
  {\tilde{V}_1^2 + \tilde{V}_2^2}
\end{eqnarray}

On the other hand, in the high temperature regime defined by $k_B T\gg
\frac{1}{\sqrt{\tilde{V}_1 \, \tilde{V}_2}}$, the conductance reaches
a saturation value,
\begin{equation}
  \label{gsat}
  G_{th}^{\rm c}= \frac{ \sqrt{2} k_B  }{\pi \hbar}
  \frac{\sqrt{\tilde{V}_1 \, \tilde{V}_2}}
  {\tilde{V}_1^2 +  \tilde{V}_2^2}, \;\;\;\;\;\;\;\;\;\;\;\;\; k_B T\gg \frac{1}{\sqrt{\tilde{V}_1 \, \tilde{V}_2}}.
\end{equation}

In Fig. \ref{fig10} we show the thermal conductance at a fixed
temperature $T$ within the regime $k_B T\gg \Delta$ as function of the
coupling strength.  As in the case of tunneling contacts, the
conductance decreases as the coupling goes to zero and as the coupling
goes to infinity, while it peaks in between. This suggests a similar
underlying mechanism to explain this behavior. The nature of the
contact is, however different and the coupling mismatching is in this
case quantified by the parameter ${\cal M}_{\alpha}= (k_B T)
\tilde{V}_{\alpha}$. A dimensional analysis indicates that
$[V_{\alpha}] = E L^2$. Then, it is appropriate to recast this
parameter as $V_{\alpha}= {\cal V}_{\alpha}/(k^{\alpha}_F k_F)$, where
$[{\cal V}_{\alpha}]= E$ and $k^{\alpha}_F$ and $k_F$ are the Fermi
wave vectors for particles with the Fermi energy within the reservoirs
$\alpha$ and the edge, respectively. With these definitions
$\tilde{V}_{\alpha}= \sqrt{\nu \nu_{\alpha}} {\cal
  V}_{\alpha}/(\varepsilon^{\alpha}_F \varepsilon_F )$, where
$\varepsilon^{\alpha}_F$ and $\varepsilon_F$ are the Fermi energy of
the electrons within the reservoirs and the edge, respectively. Thus,
$[\tilde{V}_{\alpha}]= E^{-1}$ and the mismatching measurement ${\cal
  M}_{\alpha}= (k_B T)\sqrt{\nu \nu_{\alpha}} {\cal V
}_{\alpha}/(\varepsilon^{\alpha}_F \varepsilon_F ) $ is dimensionless
and can be interpreted as a ratio between the thermal energy times the
coupling energy at the contact and the energy of the particles within
the reservoir times the energy of the particles within the edge. As
the temperature enters the matching measurement, the optimal coupling
for which the conductance achieves its maximum value depends on $T$,
as shown in Fig. \ref{fig10}. This behavior contrasts to the one of
the thermal conductance for tunneling coupling, in which case the
maximum is independent of $T$ (see Fig.  \ref{fig9}). The dependence
on $T$ of the matching measurement also suggests that the conductance
saturates at high temperature. In fact, notice that in order to
satisfy the quantum limit $G^{\rm c}_{th} \leq \kappa_0 T$,
\cite{pendry} the following condition must be fulfilled $\lambda_2
(k_B T)^2 \leq 1$, which implies a constant value of $G^{\rm c}_{th}$
for $ (k_B T)^2 > 1/(\tilde{V}_1 \tilde{V}_2)$ as shown in
Eq. \eqref{gsat}.

\begin{figure} \centering
  \includegraphics{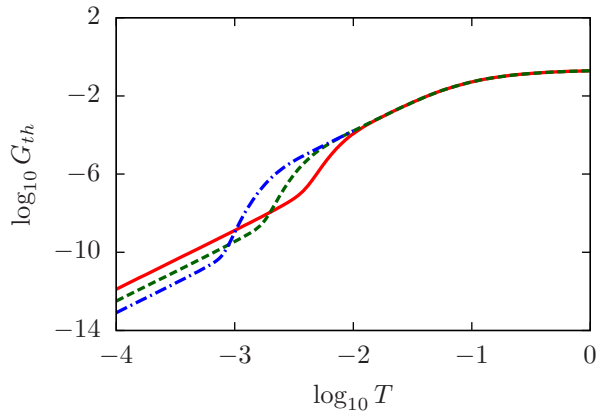}
  \caption{(Color online) Capacitive thermal conductance as a function
    of the temperature for different ring lengths: $L=100$ (red,
    solid), $200$ (green, dashed), $400$ (blue, dot-dashed)). The
    couplings are: $V_{1}=V_{2}=1$.}\label{fig6}
\end{figure}

The behavior of the thermal conductance within the different regimes
discussed in the present section are illustrated in Fig. \ref{fig6}
for systems with different lengths.  Notice that the length of the
system affects only the low temperature cubic regime $k_B T <
\Delta$. A final remarkable feature worth of notice is the dependence
on the filling factors $\nu_{\alpha}$ and $\nu$ of the thermal
conductance within the three regimes.

\subsection{Local temperature}
\subsubsection{Tunneling coupling}

In order to define the local temperature, we follow a procedure
similar to that originally introduced in Ref. \onlinecite{eng-an},
which was also the one adopted in Ref. \onlinecite{us}. We define a
thermometer consisting in a third reservoir which is non-invasively
locally coupled to the edge at a point $x_3$, i.e. $w_3 \rightarrow
0$.  The latter reservoir is assumed to satisfy the condition of local
thermal equilibrium with the edge, which means that it has a
temperature $T_3$ for which the heat current $J^{Q,{\rm t}}_3=0$. The
so determined temperature $T_3$, precisely, defines the local
temperature sensed by the thermometer.  We focus on the limit of low
temperatures, compared to the chemical potential $\mu$ but within the
range $T \gg \Delta$. The calculation is analogous to that of the
thermal conductance.
\begin{equation}
  T_{3}=
  \begin{cases}
    \sqrt{A^2 T_1^2 + B^2 T_2^2}
    & \qquad\text{if \ensuremath{x_{1}<x_{2}<x_{3}}}\\
    \sqrt{C^2 T_1^2 + D^2 T_2^2} & \qquad\text{if
      \ensuremath{x_{1}<x_{3}<x_{2}}},
  \end{cases}
\end{equation}
where the coefficients $A, B, C, D$ are given by
\begin{eqnarray}
  \label{ABCD}
  A &=& \frac{\tilde{w}_1
    \left(1-\tilde{w}^2_{2} \right)}
  {\sqrt{\left(1+\tilde{w}_{1}^{2}\tilde{w}_{2}^{2}\right)
      \left(\tilde{w}_{1}^{2}+\tilde{w}_{2}^{2}\right)}} \\
  B &=& \frac{\left(1 + \tilde{w}^2_{1} \right)
    \tilde{w}_2}
  {\sqrt{\left(1+\tilde{w}_{1}^{2}\tilde{w}_{2}^{2}\right)
      \left(\tilde{w}_{1}^{2}+\tilde{w}_{2}^{2}\right)}} \\
  C &=& \frac{\tilde{w}_1
    \left(1 + \tilde{w}_{2}^2 \right) }
  {\sqrt{\left(1+\tilde{w}_{1}^{2}\tilde{w}_{2}^{2}\right)
      \left(\tilde{w}_{1}^{2}+\tilde{w}_{2}^{2}\right)}} \\
  D &=& \frac{\left(1-\tilde{w}_{1}^2\right)
    \tilde{w}_2}
  {\sqrt{\left(1+\tilde{w}_{1}^{2}\tilde{w}_{2}^{2}\right)
      \left(\tilde{w}_{1}^{2}+\tilde{w}_{2}^{2}\right)}},
\end{eqnarray}
where again, $\tilde{w}_\alpha = w_\alpha/2\hslash \sqrt{v_F
  v_F^\alpha}$.

\begin{figure}
  \centering \subfloat[][Tunneling coupling]
  {\includegraphics{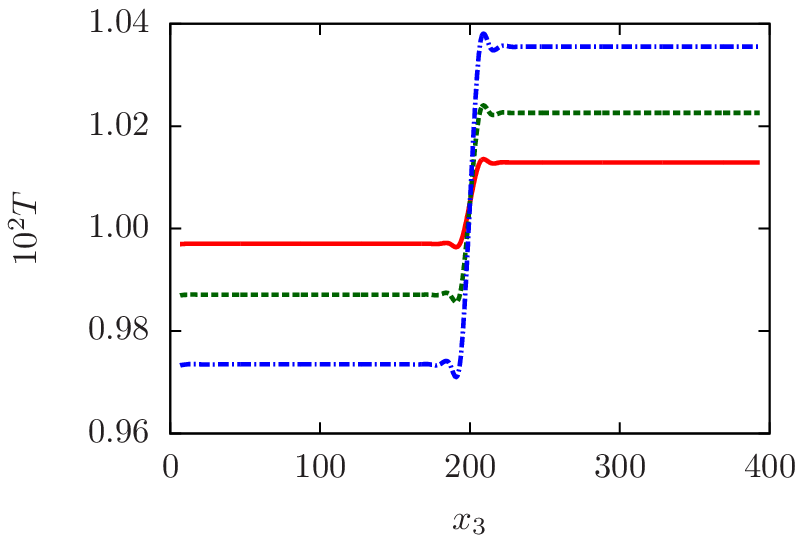}
    \label{fig7}}

  \subfloat[][Capacitive
  coupling]{\includegraphics{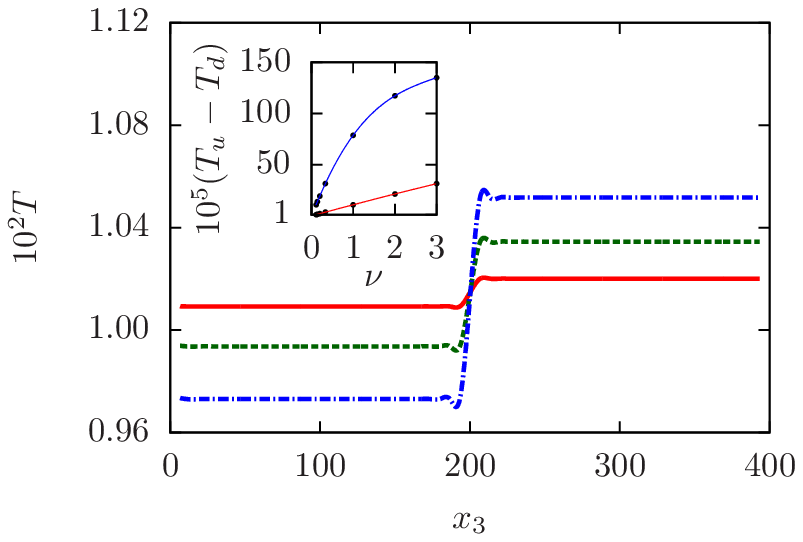}\label{fig8}}

  \caption{(Color online) Local temperature along the edge, as
    function of the position of the thermometer, for different values
    of the couplings with the reservoirs. For the tunneling case
    (Fig. \ref{fig7}) we have set $w_1=w_2=0.4$ (red, solid), $0.6$
    (green, dashed), $0.8$ (blue, dotted-dashed) and $\mu=0$. For the
    capacitive case (Fig. \ref{fig8}), the values are $V_1=V_2=1.$
    (red, solid), $2.$ (green, dashed), $3.$ (blue, dotted-dashed).
    The remaining parameters are $x_{1}=0.$, $x_{2}=200.$, $L=400.$,
    $T_{1}=0.009$, $T_{2}=0.011$.  A filling factor $\nu=1$ is
    considered. The temperature difference between the two arms of the
    ring $(T_u-T_d)$ as function of $\nu$ is shown in the inset, for
    two values of the couplings: $V_1=V_2=1., 3.$ }

\end{figure}

The results are shown in Fig.~\ref{fig7} for the choice of coupling
parameters $w_1=w_2$ between the ring and the source and drain
reservoirs, which are kept at the same chemical potential but
different temperatures $T_1> T_2$. In Fig.\ref{fig7} the profile for
the local temperature as a function of the position $x_3$ at which the
thermometer is connected is shown. The salient features of the
Fig. are the discontinuities at the positions where the source and
drain reservoirs are connected, which increase as the strength of the
couplings to the source and drain reservoirs increase. Such a behavior
is completely equivalent to that obtained in Ref.  \onlinecite{us},
where heat transport was induced by injection of an ac current at the
source reservoir, instead of establishing an explicit temperature
gradient.

The emergent picture is the following. Hot electrons tunnel from the
source reservoir and propagate along the edge with a definite
chirality along a given arm of the edge until they reach the colder
drain reservoir, to which they can tunnel. Cold electrons tunnel from
the drain reservoir and propagate with a given chirality along the
other arm of the edge until they reach the drain reservoir, to which
they can tunnel. The net result is the downstream arm of the edge
mainly visited by hot electrons, while the upstream arm is visited by
colder ones. The consequence is a higher temperature for the first arm
of the edge in comparison to the second one, as observed in the
Fig.\ref{fig7}. The temperature along the arms is approximately
constant, with small finite size oscillations of ${\cal O}(1/L)$,
which are related to the structure of levels spaced in $\Delta \propto
1/L$.  The average values of these temperatures within each arm
increases with the amplitude of the coupling to the reservoirs
$w_1,w_2$.

\subsubsection{Capacitive coupling}

In analogy to the tunneling case, we define a thermometer as a third
weakly coupled reservoir. In the present case we assume a capacitive
coupling $V_3 \rightarrow 0$ between the edge and the thermometer. As
in the previous case, we define the local temperature as the
temperature $T_3$ of the third reservoir for which there is no heat
current between this system and the edge, i.e. $J^{Q, {\rm c}}_3=0$.
The corresponding temperature profiles are shown in Fig. \ref{fig8}.
In comparison to the plots for tunneling coupling shown in
Fig. \ref{fig7}, we find the same qualitative behavior for the local
temperature along the edge.  Namely, discontinuities at the positions
where the reservoirs are coupled and thermalization within each arm
with the upstream reservoir, in agreement with the chiral propagation
of the particles along the edge state. Concerning the effect of the
filling factor on the temperature profile, its role is similar to a
renormalization of the strength of the coupling to the reservoirs
$V_1, \; V_2$, as can be inferred from the dependence of the Green
functions on $\nu$ (see Eq. \eqref{nonin23}). Thus, the effect of
thermalization of each branch with the closest upstream reservoir is
more pronounced as this parameter increases.  This is illustrated in
the inset of Fig. \ref{fig8}, where we show the difference between the
average temperature along the downstream branch $T_u$ and the
corresponding one to the upstream one $T_d$, as a function of $\nu$
for two different values of the couplings to the reservoirs. The
temperature jump $T_u-T_d$ increases with $\nu$ following a
non-universal law.

\section{Summary and conclusions}

We have analyzed the heat transport through edge states of a
two-dimensional electron gas in the quantum Hall effect. We considered
two exactly solvable configurations. One of the cases corresponds to a
system with filling factor $\nu=1$ coupled to reservoirs at different
temperatures through tunneling couplings. The second case corresponds
to capacitive coupling between the edge and the reservoirs and integer
or fractional filling factor of the form $\nu=1/m$.

The main features can be characterized in terms of two temperature
regimes, defined with respect to the level spacing of the edge,
$\Delta = \frac{2 \pi \hbar v_F}{L}$.  In the mesoscopic regime, $T
\ll \Delta$, finite size effects related to the discrete level spacing
of the edge affect the behavior of the heat transport.  In the
tunneling case, the two terminal thermal conductance $G^{\rm t}_{th}$
has a different behavior depending on the position of the chemical
potential, relative to the positions of the energy levels of the
edge. For ultra-low temperatures, $T\ll \gamma$, where $\gamma$ is the
width of the peaks, it grows linearly with $T$, obeying the universal
law $G^{\rm t}_{th}=\kappa_0\,T$, with $\kappa_0$ the universal
thermal conductance quantum.

In the capacitive case, the thermal conductance behaves in a very
different way within this regime. It displays a highly non-universal
cubic law, $G^{\rm c}_{th} \propto T^3$, with a coefficient depending
on the couplings to the reservoirs and the edge length, through the
combination $\frac{V_1^2\,V_2^2}{L^2}$.

In the macroscopic regime, which takes place at higher temperatures,
$T > \Delta$, finite-size effects become irrelevant, and thermal
transport does not depend on the length of the system.  However, the
regime is not universal, in the sense that there is a strong
dependence on the couplings to the reservoirs, for both tunneling and
capacitive contacts.  In particular, in the tunneling case, the two
terminal thermal conductance is still a linear function of
temperature, but with a proportionality coefficient that is a
non-monotonic function of the coupling strengths (see
Eq. \eqref{fw1w2}). In the capacitive case, this regime extends to any
other filling $\nu$.  Interestingly, an additional intermediate regime
appears in this case for $\Delta \ll k_BT \ll
\frac{1}{\sqrt{\tilde{V}_1\,\tilde{V}_2}}$, with a thermal conductance
characterized again by a cubic power law behavior, $G^{\rm c}_{th}(T)=
\kappa T^3$, with $\kappa$ depending on the filling factors $\nu$,
$\nu_1$, $\nu_2$, and the couplings $V_1$ and $V_2$.  As in the
tunneling case, $\kappa$ is a non-monotonic function of the couplings
(see Eq. (\ref{kappa-cap2})).  Finally, at high temperatures, $k_B T
\gg \frac{1}{\sqrt{\tilde{V}_1\,\tilde{V}_2}}$, the capacitive thermal
conductance reaches a saturation value, $G^{\rm c}_{th}\sim
\frac{1}{\sqrt{\nu}}\,\frac{\sqrt{\sqrt{\nu_1 \, \nu_2}\, \tilde{V}_1
    \, \tilde{V}_2}}{\nu_1 \, \tilde{V}_1^2 + \nu_2 \,
  \tilde{V}_2^2}$.  This result strongly differs from the linear
behavior of the tunneling thermal conductance in the same regime.
Concerning the dependence on the filling factors, we are not able to
exactly treat the case with $\nu \neq 1$ for tunneling contacts, in
order to verify if the conductance is also independent of the
filling. However, if we notice that the two terminal electrical
conductance for point-like tunneling contacts is $G=e^2/h$,
independent of $\nu$,\cite{cham-frad} it is likely that this
independence also holds for the thermal conductance in the presence of
tunneling couplings.

Our results indicate that the behavior of the local temperature along
the edge, defined from the coupling to a thermometer, is qualitatively
the same for tunneling and capacitive couplings. As in the case of the
ac-driven edge considered in Refs. [\onlinecite{granger,us}], the
landscape of local temperature as a function of the position along the
edge, is fully consistent with the chiral propagation of the energy
along the edge state. This means that given a configuration of an edge
in contact to reservoirs with different temperatures, each piece of
the edge tends to thermalize with the closest upstream
reservoir. Remarkably, this behavior does not depend on the nature of
the contact. It is qualitatively the same for a tunneling contact,
which injects particles that carry energy, and for capacitive contact
where only energy is exchanged.

The practical outcome of our results is the fact that capacitive
couplings are as suitable as tunneling ones for the study of heat
along edge states. This opens the possibility for the study of hybrid
setups including capacitive and tunneling connections. This could be
particularly interesting in the studies of edge states of quantum Hall
states with fillings $\nu=2/3$ and $\nu=5/2$ that have charged along
with neutral modes, which are insensitive to a capacitive coupling.
On the other hand, since our results indicate that thermal transport
with capacitive contacts is sensitive to the values of $\nu$, $\nu_1$
and $\nu_2$, one could conceive a capacitive thermal device designed
to measure filling fractions of quantum Hall samples.


\begin{acknowledgments}
  We thank G. Lozano for discussions. We acknowledge support from
  CONICET, ANPCyT, UBACYT, UNLP (Argentina). EF thanks Programa
  Ra{\'\i}ces (MINCYT, Argentina) for support and the Department of
  Physics, FCEyN UBA (Argentina) for hospitality. This work was
  supported in part by the National Science Foundation, under grant
   DMR-1064319 (EF).
\end{acknowledgments}

\appendix
\section{Dyson equations for the fermionic Green functions. Tunneling
  coupling}
\label{apa}
We define the mixed retarded Green function
\begin{equation}
  \Gmixed^R(x, r_{\alpha}; t, t^{\prime}) =
  - i \Theta(t-t^{\prime})
  \langle
  \{
  \Psi(x,t) , \Psi^{\dagger}(r_\alpha,t^{\prime} )
  \}
  \rangle,
\end{equation}
and the retarded Green function of the ring
\begin{equation}
  G^R(x, x^\prime; t, t^\prime) =
  - i \Theta(t-t^\prime)
  \left\langle
    \{ \Psi(x,t) , \Psi^{\dagger}(x^\prime, t^{\prime} )\}
  \right\rangle
\end{equation}
The Dyson equations for this functions cast
\begin{widetext}
  \begin{subequations}
    \begin{eqnarray} \label{dys1} \left( \partial_{t^{\prime}}-
        v_F\partial_{x^{\prime}} \right) G^R(x,x^{\prime};t -
      t^{\prime}) &=& \delta(t-t^{\prime}) \delta(x-x^{\prime}) +
      \sum_{\alpha} w_{\alpha} 
      \Gmixed^R(x,r_{\alpha};t - t^{\prime}) \delta(r_{\alpha}-r_{\alpha}^0) \delta(x-x_{\alpha})  \\
      \left( \partial_{t^{\prime}}- v_F^{\alpha}\partial_{r_{\alpha}}
      \right) \Gmixed^R(x,r_{\alpha};t - t^{\prime})&=& w_{\alpha}
      G^R(x,x^{\prime};t - t^{\prime}) \delta(r_{\alpha}-r_{\alpha}^0)
      \delta(x^{\prime}-x_{\alpha}).
    \end{eqnarray}
  \end{subequations}
\end{widetext}
We introduce the inverse of the differential operators
\begin{eqnarray}
  \label{invop}
  \left( \partial_{t^{\prime}}- v_F\partial_{x^{\prime}} \right)
  G^{0,R}(x - x^{\prime};t - t^{\prime}) &=&
  \delta(t - t^{\prime}) \\
  &\times&
  \delta(x -x^{\prime}) \nonumber \\
  \left( \partial_{t^{\prime}}- v_F^{\alpha}\partial_{r_{\alpha}} \right)
  g^R(r_\alpha - r_\alpha^\prime, t - t^\prime)
  &=& \delta(t - t^{\prime}) \\
  &\times&
  \delta(r_\alpha - r_\alpha^{\prime}) \nonumber,
\end{eqnarray}
which we identify as the \emph{free} retarded Green functions of the
ring and the reservoirs, respectively.  Performing the Fourier
transform with respect to $t-t^\prime$ and $r_\alpha -
r_\alpha^\prime$ in the Eqs. \eqref{dys1} and \eqref{invop} we obtain
\begin{subequations}
  \begin{eqnarray}
    \label{dya}
    G^R(x,x^{\prime};\omega) & = & G^{0, R}(x,x^{\prime};\omega) \nonumber \\
    & - & \sum_{p_{\alpha}} {\tilde G}^R(x,p_{\alpha};\omega)   \\
    & & \qquad \times w_{p_{\alpha}}
    G^{0, R}(x_{\alpha},x^{\prime};\omega) ,  \nonumber \\
    \label{dyb}
    \Gmixed^R(x,p_{\alpha};\omega) & = & - G^R(x,x_{\alpha};\omega)
    w_{p_{\alpha}} g^R(p_{\alpha}, \omega),
  \end{eqnarray}
\end{subequations}
where $w_{p_{\alpha}}=w_{\alpha} e^{-i p_{\alpha}
  r^0_{\alpha}}/\sqrt{L_{\alpha}}$ and
\begin{equation}
  \label{eq:3}
  g^R(p_{\alpha}, \omega)
  =  [g^A(p_{\alpha},\omega)]^*= \frac{1}{\omega -
    \varepsilon_{p_{\alpha}} + i \eta},
\end{equation}
with $\varepsilon_{p_{\alpha}}= v_F^{\alpha} p_{\alpha}$. The equation
\eqref{dyb} can be replaced in \eqref{dya} obtaining
\begin{equation}
  \label{dyret-tun}
  \begin{split}
    G^R(x,x^{\prime};\omega) &= G^{0,R}(x,x^{\prime};\omega) \\
    & + \sum_{\alpha=1}^M G^R(x,x_{\alpha};\omega)
    \Sigma^{t,R}_{\alpha}(\omega)
    G^{0,R}(x_{\alpha},x^{\prime};\omega),
  \end{split}
\end{equation}
where we have defined the retarded self-energies
\begin{equation}
  \Sigma^{t,R}_{\alpha}(\omega) = |w_{\alpha}|^2 \int \frac{dp_{\alpha}}{2 \pi} g^R(p_{\alpha},\omega)
\end{equation}
which enclose the effect of the coupling to the reservoir.  It is
useful to define the spectral function
\begin{equation}
  \Gamma^t_{\alpha}(\omega) = - 2
  \mbox{Im}[\Sigma^{t,R}_{\alpha}(\omega)]
\end{equation}
which explicitly reads
\begin{equation}
  \label{gammat}
  \Gamma^t_{\alpha}(\omega) = \frac{|w_{\alpha}|^2}{v_F^{\alpha}}
  \Theta (\Lambda-|\omega|)
\end{equation}
where $\Lambda$ is a high energy cutoff which defines the bandwidth of
the reservoir, while the corresponding explicit expression for the
retarded self-energy is
\begin{equation}
  \label{self-tun}
  \Sigma^{t,R}_{\alpha}(\omega) = \mbox{ln}\left|\frac{\Lambda-
      \omega}{\Lambda+ \omega}\right|- i
  \frac{\Gamma_{\alpha}^t(\omega)}{2} .
\end{equation}
It is easy to verify that the Green functions evaluated from
(\ref{dyret-tun}) satisfy the following identity
\begin{equation}
  \label{idt}
  \begin{split}
    G^R(x,x^{\prime};\omega) &- G^A(x^{\prime}, x;\omega) = - i
    \sum_{\beta=1}^M
    G^R(x,x_{\beta};\omega) \nonumber \\
    & \times \Gamma^{t}_{\beta}(\omega) G^A(x_{\beta},
    x^{\prime};\omega),
  \end{split}
\end{equation}
being $G^A(x^{\prime}, x; \omega)=[G^R(x,x^{\prime};\omega) ]^*$ the
advanced Green function.

In order to calculate the current, we need the lesser mixed Green
function.  The Dyson's equations can be straightforwardly derived from
Eq.\eqref{dyret-tun} using Langreth's rules\cite{lang}
\begin{eqnarray}
  \label{dyles-tun}
  \Gmixed^<(x,p_{\alpha};\omega) & = & - G^<(x,x_{\alpha};\omega)
  w_{p_{\alpha}}
  g^A(p_{\alpha},\omega) \nonumber \\
  & & - G^R(x,x_{\alpha};\omega) w_{p_{\alpha}}
  g^<(p_{\alpha},\omega),\nonumber \\
  G^<(x,x^{\prime};\omega) & = & \sum_{\alpha=1}^M
  G^R(x,x_{\alpha};\omega) \Sigma^{t,<}_{\alpha}(\omega) \nonumber\\
  &\times&[G^R(x^{\prime},x_{\alpha};\omega)]^*
\end{eqnarray}
being
\begin{equation}
  \Sigma_{\alpha}^{t,<}(\omega) =
  i \int \frac{d \omega}{2 \pi}
  f_{\alpha}(\omega) \Gamma^t_{\alpha}(\omega),
\end{equation}
while
\begin{equation}
  g^<_{p_{\alpha}}(\omega) =  2 i \pi
  f_{\alpha}(\omega) \delta(\omega - \varepsilon(p_{\alpha}))
\end{equation}
with
\begin{equation}
  f_{\alpha}(\omega)=\frac{1}{e^{(\omega-\mu)/T_{\alpha}}+1}
\end{equation}
being the Fermi-Dirac distribution.

The Green function of the decoupled ring is given by the sum
\begin{equation}
  \label{free-ring}
  G^{0, R}(x,x^{\prime};\omega)= \frac{1}{L} \sum_{n=-N}^N \frac{e^{i
      k_n(x-x^{\prime})}}{\omega - \varepsilon_{k_n} + i \eta},
\end{equation}
with $\varepsilon_k = v_F k$, being $k_n= 2 n \pi /L$ and $N$ is a
positive integer which defines the high energy cutoff $\Lambda_r= v_F
2 \pi N /L$.  In the limit of $N \rightarrow \infty$ the sum in the
above non-interacting Green function can be computed
analytically\cite{GR}, and the result is
\begin{equation}
  \begin{split}
    \label{nonin22}
    & G^{0, R}(x,x^{\prime};\omega) = \frac{1}{2 v_F} \frac{e^{-i
        \omega(x-x^{\prime})/v_F }}
    {\sin(\omega L/2 v_F)}  \\
    & \times \{ \Theta(x-x^{\prime}) e^{i \omega L/(2v_F)} +
    \Theta(x^{\prime}-x) e^{-i \omega L/(2v_F)} \}
  \end{split}
\end{equation}
This last expression has been extensively used in all the analytical
calculations.

\section{Dyson's equation for the bosonic Green function. Capacitive
  coupling}
\label{apb}

We follow a similar procedure to that exposed in the previous
appendix. The differences are just due to the different type of
commutation relations of Eq.\eqref{boscom} obeyed by the chiral
bosonic fields.  As in the fermionic case, we start by defining the
retarded Green function corresponding to the ring
\begin{equation}
  D^R(x, x^\prime;t, t^{\prime})  =
  -i \Theta(t-t^{\prime})
  \langle [\phi(x,t) , \phi(x^\prime, t^{\prime}) ]
  \rangle
\end{equation}
and mixed degrees of freedom \begin{equation} {\tilde
    D}^R(x,r_\alpha;t, t^{\prime}) = -i \Theta(t-t^{\prime}) \langle
  [\phi(x,t) , \phi(r_{\alpha}, t^{\prime}) ] \rangle.
\end{equation}
We derive the Dyson's equations by evaluating the time evolution of
the fields. For the retarded Green functions defined above we get
\begin{widetext}
  \begin{subequations}
    \begin{align}
      \label{dys1a}
      \frac{1}{\nu} \left(
        \partial_{t^{\prime}}- v_F\partial_{x^{\prime}} \right)
      D^R(x,x^{\prime};t - t^{\prime}) &=
      \, \pi \, \delta(t-t^{\prime}) \sgn(x-x^{\prime}) \nonumber \\
      &+ 2 \pi \sum_{\alpha} V_{\alpha}
      \partial_{r_{\alpha}} {\tilde D}^R(x,r_{\alpha};t - t^{\prime})
      \delta(r_{\alpha}-r_{\alpha}^0) \delta(x-x_{\alpha})\\
      \label{dys1b}
      \frac{1}{\nu_\alpha} \left(
        \partial_{t^{\prime}}- v_F^{\alpha}\partial_{r_{\alpha}}
      \right) \tilde{D}^R(x,r_{\alpha};t - t^{\prime}) &= 2 \pi
      V_{\alpha}
      \partial_{x^{\prime}} D^R(x,x^{\prime};t - t^{\prime})
      \delta(r_{\alpha}-r_{\alpha}^0) \delta(x^{\prime}-x_{\alpha}).
    \end{align}
  \end{subequations}
\end{widetext}
It is now convenient to introduce the inverse of the operators
$\left(\partial_{t^{\prime}}-
  v_F^{\alpha}\partial_{r_{\alpha}^{\prime}}\right) $ and
$\left(\partial_{t^{\prime}}- v_F^{\alpha}\partial_{x^{\prime}}\right)
$, which correspond to the solutions of the following partial
differential equations
\begin{eqnarray}
  \label{cap-diff1}
  \frac{1}{\nu_\alpha}
  \left(
    \partial_{t^{\prime}}- v_F^{\alpha}\partial_{r_{\alpha}^{\prime}}
  \right)
  d^{0, R}(r_{\alpha},r^{\prime}_{\alpha}; t - t^{\prime}) &=&
  \delta(t-t^{\prime}) \\
  &\times&  \delta(r_{\alpha}-r^{\prime}_{\alpha}) \nonumber \\
  \label{cap-diff2}
  \frac{1}{\nu}\{ \partial_{t^{\prime}}- v_F\partial_{x^{\prime}} \}
  D^{0, R}(x,x^{\prime};t - t^{\prime}) &=& \delta(t-t^{\prime}) \\
  &\times& \delta(x-x^{\prime}). \nonumber
\end{eqnarray}
The Fourier transform of these functions read
\begin{eqnarray}
  \label{eq:13}
  d^{0, R}(r_{\alpha},r_{\alpha}^{\prime};\omega) &=&
  \nu_\alpha
  \int \frac{d p_{\alpha}}{2 \pi}
  \frac{e^{-i p_{\alpha} (r_{\alpha} - r_{\alpha}^{\prime} )}}
  {i( \omega - v_F^{\alpha}p_{\alpha} + i \eta )}, \\
  \label{nonin23}
  D^{0, R}(x,x^{\prime};\omega) &=& \frac{\nu}{L} \sum_{n=-N}^N \frac{e^{i
      k_n(x-x^{\prime})}}{\omega - \varepsilon_{k_n} + i \eta}.
\end{eqnarray}
Fourier transforming with respect to $t-t^\prime$, combining the
equations \eqref{dys1a} and \eqref{dys1b}, and calculating the
derivatives with respect to $x, x^{\prime}$ results in the expression
\begin{eqnarray}
  \label{dys4}
  \begin{split}
    {\cal D}^{R}(x,x^{\prime};\omega) &= {\cal
      D}^{0, R}(x,x^{\prime};\omega) \\
    &+ \sum_{\alpha=1}^M{\cal D}^{R}(x,x_{\alpha};\omega) \times
    \Sigma^{c,R}_{\alpha}(\omega) {\cal D}^{0,
      R}(x_{\alpha},x^{\prime};\omega),
  \end{split}
\end{eqnarray}
where
\begin{equation}
  \label{cald}
  {\cal D}^{R}(x,x^{\prime};\omega)=\frac{1}{2\pi}  \partial_x
  \partial_{x^{\prime}} D^R(x,x^{\prime};\omega)
\end{equation}
and
\begin{equation} {\cal D}^{0, R}(x,x^{\prime};\omega)
  =\partial_{x^{\prime}} D^{0, R}(x,x^{\prime};\omega)
\end{equation}
Thus, as in the tunneling case, we have eliminated the degrees of
freedom of the reservoirs from the Dyson's equation by defining
self-energies
\begin{align}
  \Sigma^{c,R}_{\alpha}(\omega) =& (2 \pi V_{\alpha})^2
  \frac{\nu_\alpha}{v_F^{\alpha}}\nonumber\\
  \times & \lim_{r_{\alpha}^{\prime} \rightarrow {r_{\alpha}^0}^+}
  \partial_{r^{\prime}_{\alpha} } \int_{-P}^{+P} \frac{d p_{\alpha}}{2
    \pi i} \frac{ e^{-i p_{\alpha}(r^0_{\alpha}- r_{\alpha}^{\prime}
      )}}{\omega - v_F^{\alpha} p_{\alpha} + i \eta} ,
\end{align}
which depend on the coupling to the reservoir as well as on their
density of states. In the limit of the cutoff $\Lambda = v_F P
\rightarrow \infty$ we obtain
\begin{equation}
  \label{sigma-explicit}
  \Sigma_{\alpha}^{c,R}= \frac{i}{2} \; \Gamma_{\alpha}^c(\omega),
\end{equation}
with \be
\label{gammac}
\Gamma_{\alpha}^c(\omega) = (2\pi)^2
\frac{V_{\alpha}^2}{(v_F^{\alpha})^2} \nu_{\alpha} \omega \;
\Theta(\Lambda-\mid\omega\mid).  \ee

Notice that Eq. \eqref{dys4} has the same structure as
\eqref{dyret-tun}. These Green functions, thus, satisfy the following
identity, analogous to \eqref{idt}, \ba
\label{equal} {\cal D}^{R}(x,x^{\prime};\omega)& - & {\cal
  D}^{A}(x^{\prime},x;\omega)=- i \sum_{\alpha=1}^M {\cal
  D}^{R}(x,x_{\alpha};\omega) \nonumber \\
& & \times \Gamma^c_{\alpha}(\omega) {\cal D}^{A}(x_{\alpha},
x^{\prime};\omega) , \ea with ${\cal D}^{A}(x^{\prime},x;\omega)=
[{\cal D}^{R}(x,x^{\prime};\omega)]^*$.

From \eqref{dys4} we can derive the Dyson equations for the lesser
Green functions by using the Langreth rules\cite{lang}
\begin{eqnarray}
  \label{dys5}
  {\cal D}^{<}(x,x^{\prime} ;\omega)& = &
  \sum_{\alpha=1}^M {\cal
    D}^R(x,x_{\alpha};\omega) \Sigma^{c,<}_{\alpha}(\omega)  \nonumber \\
  & & \times {\cal
    D}^{A}(x_{\alpha},x^{\prime};\omega) ,
\end{eqnarray}
with
\begin{equation}
  \Sigma^{c,<}_{\alpha}(\omega)= i \Gamma^c_{\alpha}(\omega)
  n_{\alpha}(\omega)
\end{equation}
where
\begin{equation}
  n_{\alpha}(\omega)=\frac{1}{e^{\omega/T_{\alpha}}-1}
\end{equation}
is the Bose-Einstein distribution function corresponding to the
temperature $T_{\alpha}$ of the reservoir.

Finally, using Langreth rules in the Fourier transform of
Eq. \eqref{dys1b} and performing the derivative with respect to $x$
and the second derivative with respect to $r_{\alpha}$ we obtain
\begin{widetext}
  \ba
  \partial_x \partial^2_{r_{\alpha}} D^<(x,
  r^{\prime}_{\alpha};\omega) |_{r_{\alpha}=r_{\alpha}^0} & = & (2
  \pi)^2 V_{\alpha} \{ {\cal D}^<(x,x^{\prime};
  \omega)|_{x^{\prime}=x_{\alpha}}
  \partial^2_{r_{\alpha}} D^{0,A}(r_{\alpha}^0,r_{\alpha};\omega)
  |_{r_{\alpha}=r_{\alpha}^0} \nonumber \\
  & & + {\cal D}^R(x,x^{\prime}; \omega)|_{x_{\alpha}}
  \partial^2_{r_{\alpha}} D^{0,<}(r_{\alpha}^0,r_{\alpha};\omega)
  |_{r_{\alpha}=r_{\alpha}^0} \}, \ea
\end{widetext}
where $D^{0,A}(r_{\alpha}, r_{\alpha}^{\prime};\omega)=
[D^{0,R}(r_{\alpha}^{\prime}, r_{\alpha};\omega)]^*$ and
\begin{widetext}
  \begin{equation} \label{der2}
    \partial^2_{r_{\alpha}} D^{0,<}(r_{\alpha}^0, r_{\alpha};\omega) |_{r_{\alpha}=r_{\alpha}^0}
    = - \int_{-P}^{+P} dp_{\alpha} p_{\alpha}^2   n_{\alpha}(\omega) \delta(\omega
    - v_F^{\alpha} p_{\alpha}).
  \end{equation}
\end{widetext}


\section{Analytic calculation of the thermal conductance for the
  tunneling case in the macroscopic regime}
\label{app:analytic}

We evaluate the thermal conductance for the tunneling case, in the
macroscopic regime. We will employ the method of contour
integration. In the following, we set $v_F =v_F^\alpha=1$.  Starting
from \eqref{jqtun}, and taking into account that $T=\frac{1}{\beta}\gg
\Delta$, we find that $G_{th}^t$ is given by
\begin{equation}
  G_{th}^t = \int_{-\infty}^{\infty} \frac{d\omega}{2\pi}
  \mathcal{T}^t_{12}(\omega) \frac{(\omega-\mu)^{2} \
    \beta^{2} e^{\beta(\omega-\mu)}}
  {\left(1+e^{\beta(\omega-\mu)}\right)^{2}}
  \label{integral-for-conductance},
\end{equation}
with
\begin{equation}
  \mathcal{T}^t_{12}(\omega)=\Gamma_{1}(\omega) \Gamma_{2}(\omega)\left|G^{R}(x_{1},x_{2};\omega)\right|^{2},
\end{equation}
where $\Gamma_{1}(\omega)$ is given by Eq.\eqref{gammat}, and
$G^{R}(x_{1},x_{2};\omega)$ is the full Green function of the ring.

In the limit $\Lambda \to \infty$, one has
\begin{equation}
  \begin{split}
    \mathcal{T}^t_{12}(\omega) &= \frac{w_{1}^{2}w_{2}^{2}}
    {4\left(1+w_{1}^{2}w_{2}^{2}/16\right)^{2}} \\
    &\quad \times \frac{1} {\sin^{2} \left(\omega L/2 \right) + \left(
        \epsilon L/2 \right )^{2} \cos^{2} \left(\omega L/2 \right) },
  \end{split}
\end{equation}
where we defined
\begin{displaymath}
  \frac{\epsilon L}{2} \doteq
  \frac{(w_{1}^{2}+w_{2}^{2})}{4\left(1+w_{1}^{2}w_{2}^{2}/16\right)}.
\end{displaymath}
We evaluate the integral in Eq.\eqref{integral-for-conductance} by
closing the contour over the upper half complex plane. The poles that
lie inside the contour are located at the points
\begin{equation}
  \omega_{n}^{F} = \mu + \frac{\left(2n+1\right)\pi i}{\beta},
  \qquad n=0,1,2,\ldots
\end{equation}
and
\begin{equation}
  \omega_{n}  = \frac{i}{L}\ln\left|\frac{1+\epsilon L/2}{1-\epsilon L/2}\right|+\frac{2n\pi}{L}
\end{equation}
The first series of poles does not contribute in the limit $T \gg
\Delta$. The residues of the second series of poles are given by
\[
\begin{split}
  \text{Res}(\omega_{n}) &= \frac{-i}{2\pi} \frac{w_{1}^{2}w_{2}^{2}}
  {4\left(1+w_{1}^{2}w_{2}^{2}/16\right)^{2}} \frac{2}{L}
  \frac{\sqrt{1-\left(\epsilon L/2\right)^{2}}}
  {\epsilon L/2} \\
  &\times \frac{\beta^{2} \left(2n\pi/L- \mu \right)^{2}
    e^{\beta\left(2n\pi/L-\mu\right)}}
  {\left(1+e^{\beta\left(2n\pi/L-\mu\right)}\right)^{2}}.
\end{split}
\]
The integral can be written as a sum
\begin{equation}
  G_{th}^t = 2\pi i \sum_n \text{Res}(\omega_{n}).
\end{equation}
In the high temperature limit ($\beta/L\to0$), we can transform the
summation into an integral
\begin{equation}
  G_{th}^t = f(w_{1},w_{2}) \frac{2\pi}{L} \int_{-\infty}^{\infty}
  dn\, \frac{\left(2n\beta\pi/L-\beta\mu\right)^{2}e^{\beta\left(2n\pi/L-\mu\right)}}{\left(1+e^{\beta\left(2n\pi/L-\mu\right)}\right)^{2}}.
\end{equation}
Performing the substitution $\frac{2n\pi\beta}{L}-\beta\mu=k,$
$dk=\frac{2\pi\beta}{L}\, dn$ and evaluating the remaining integral,
we obtain
\begin{equation}
  G_{th}^t = f(w_{1},w_{2})\, \frac{\pi^{2}}{3} \,T
\end{equation}
where
\begin{equation}
  f(w_{1},w_{2}) = \frac{1}{2\pi}\frac{w_{1}^{2}w_{2}^{2}}
  {w_{1}^{2}+w_{2}^{2}}
  \frac{1}
  {\left(1+w_{1}^{2}w_{2}^{2}/16\right)}.
\end{equation}

The behavior displayed in Fig. \ref{fig9} corresponds to the case
$w_{1}=w_{2}=w$, which yields
\begin{equation}
  G_{th}^t= \frac{w^{2}}
  {4\pi\left(1+w^{4}/16\right)}
  \frac{\pi^{2}}{3} T.
\end{equation}

\section{Analytic calculation of the thermal conductance for the
  capacitive case}
\label{app:analytic-capacitive}

As in the tunneling case, we evaluate the thermal conductance by the
method of contour integration. For simplicity we set $v_F =
v_F^\alpha=1$ and $\nu =\nu_1 = \nu_2 =1$.  The dependence on these
quantities will be recovered at the end of the computation.  In the
macroscopic regime, $T=\frac{1}{\beta}\gg \Delta$, the expression of
Eq.\eqref{jqcap} leads to
\begin{equation}
  G_{th}^{c} = \int_{-\infty}^{\infty} \frac{d\omega}{2\pi} \mathcal{T}_{12}^{c}(\omega)
  \frac{\omega^{2}\beta^{2}e^{\beta\omega}}
  {\left(1-e^{\beta\omega}\right)^{2}}.
\end{equation}
In the limit $\Lambda \to \infty$, the transmission coefficient is
\begin{equation}
  \begin{split}
    \mathcal{T}_{12}^{c}(\omega) &= \frac{\left(2\pi\right)^{4}
      V_{1}^{2}V_{2}^{2}\omega^{4}} {\left(1+\pi^{4} V_{1}^{2}
        V_{2}^{2} \omega^{4}
      \right)^{2}} \\
    &\quad \times \frac{1}{4\left[\sin^{2}\left(\omega L/2\right)+
        F^2(\omega) \cos^{2}\left(\omega L/2\right)\right]},
  \end{split}
\end{equation}
where we have defined
\begin{equation}
  F(\omega) = \frac{\pi^{2}\left(V_{1}^{2}+V_{2}^{2}\right)\omega^{2}}
  {1+\pi^{4}V_{1}^{2}V_{2}^{2}\omega^{4}}.
\end{equation}
We evaluate the integral by closing the contour over the upper half
complex plane. The poles that lie inside the contour are located at
the points
\begin{equation}
  \omega_n^B = \frac{2 n \pi i}{\beta},\qquad n=1,2,\ldots
\end{equation}
and at the points given by the solutions of the transcendental
equation
\begin{equation}
  \tan \left(\frac{\omega L}{2}\right) =
  \frac{i\pi^{2}(V_{1}^{2}+V_{2}^{2})\omega^{2}}
  {1+\pi^{4}V_{1}^{2}V_{2}^{2}\omega^{4}}.
\end{equation}
This equation cannot be solved exactly. However, the solutions can be
very well approximated by
\begin{equation}
  \omega_{n} \approx \frac{2n\pi}{L}+\frac{i}{L}\ln\left[\frac{1+F(2n\pi/L)}{1-F(2n\pi/L)}\right].
\end{equation}
As in the capacitive case, the poles $\omega_n^B$ do not contribute to
the integral in the limit $\beta/L \to 0$. Thus, the integral can be
written in terms of the residues at $\omega = \omega_n$ as
\begin{equation}
  \label{eq:2}
  G_{th}^c  =  \frac{2\pi}{L}
  \frac{V_{1}^{2}V_{2}^{2}}
  {(V_{1}^{2}+V_{2}^{2})}
  \sum_{n} \frac{\omega_{n}^{4}}
  {\left(1+\pi^{4}V_{1}^{2}V_{2}^{2}\omega_{n}^{4}\right)}
  \,\frac{\beta^{2}e^{2n\pi\beta/L}}
  {\left(1-e^{2n\pi\beta/L}\right)^{2}}.
\end{equation}
In the limit $\beta/L \to 0$ we transform the summation into an
integral and we then find
\begin{equation}
  G_{th}^c=\frac{4}{\pi^2(V_{1}^{2}+V_{2}^{2})}\int_{-\infty}^{\infty}dx\frac{\pi^{4}V_{1}^{2}V_{2}^{2}x^{4}}{\left(1+\pi^{4}V_{1}^{2}V_{2}^{2}x^{4}\right)}\,\frac{\beta^{2}e^{\beta x}}{\left(1-e^{\beta x}\right)^{2}},
\end{equation}
with $x=\frac{2\pi n}{L}$.  From this integral we obtain the following
limiting behaviors, within the macroscopic regime,
$T=\frac{1}{\beta}\gg \Delta$.  At low temperatures, $T
\sqrt{V_{1}V_{2}} \ll 1$, the conductance $G_{th}^c$ exhibits a power
law behavior
\begin{equation}
  G_{th}^c = \frac{16 \pi^5}{15}  \frac{V_{1}^2 V_{2}^2}{V_{1}^{2}+V_{2}^{2}} T^3.
\end{equation}
whereas at high temperatures, $T \sqrt{V_{1}V_{2}} \gg 1$, it
approaches a finite constant value
\begin{equation}
  G_{th}^c = \frac{\sqrt{2}}{\pi} \frac{\sqrt{V_{1}V_{2}}}{(V_{1}^{2}+V_{2}^{2})}.
\end{equation}
The dependence on the filling fractions and Fermi velocities are
easily recovered by the substitutions $V_{\alpha}\rightarrow \sqrt{\nu
  \nu_{\alpha}} V_{\alpha}/ v_F v_F^{(\alpha)}$, with $\alpha=1,2$.


\begin{thebibliography}{99}
\bibitem{QH-edges} For a detailed discussion of edge states of quantum
  Hall systems see: X. -G. Wen, Adv. Phys.  \textbf{44}, 405 (1995);
  A. M. Chang, Rev. Mod. Phys. \textbf{75}, 1449 (2003), and
  E. Fradkin, {\sl Field Theories of Condensed Matter Systems}, 2nd
  edition, Cambridge University Press, (Cambridge, UK, 2013).


\bibitem{laug}R. B. Laughlin, Phys. Rev. B {\bf 23}, 5632 (1981).

\bibitem{Halperin-1982} B. I. Halperin, Phys. Rev. B \textbf{25}, 2185
  (1982).

\bibitem{but}M. B\"uttiker, Phys. Rev. B {\bf 38}, 9375 (1988).

\bibitem{wen-niu} X. G. Wen and Q. Niu, Phys. Rev. B \textbf{41}, 9377
  (1990).

\bibitem{Wen-1990} X. -G. Wen, Phys. Rev. Lett. \textbf{64}, 2206
  (1990)

\bibitem{Wen-1990b} X. G. Wen, Phys. Rev. B \textbf{41}, 12838 (1990).

\bibitem{kane-fish} C. L. Kane and Matthew P. A. Fisher,
  Phys. Rev. Lett.  {\bf 68}, 1220 (1992); Phys. Rev. B {\bf 46},
  15233 (1992); Phys.  Rev. Lett. {\bf 72}, 724 (1994).



\bibitem{granger} G. Granger, J. P. Eisenstein and J. L. Reno,
  Phys. Rev. Lett. {\bf 102}, 086803 (2009).

\bibitem{bid} A. Bid, N. Ofek, H. Inoue, M. Heiblum, C. L. Kane,
  V. Umansky and D. Mahalu, Nature \textbf{466}, 585 (2010).

\bibitem{altimiras} C. Altimiras, H. le Sueur, U. Gennser, A. Cavanna,
  D. Mailly and F. Pierre, Phys. Rev. Lett. \textbf{105}, 226804
  (2010).

\bibitem{yacoby}V. Venkatachalam, S. Hart, L. Pfeiffer, K. West, and
  A. Yacoby, Nature Physics {\bf 8}, 676 (2012).

\bibitem{altimiras2}C. Altimiras, H. le Sueur, U. Gennser, A. Anthore,
  A. Cavanna, D. Mailly, and F. Pierre, Phys. Rev. Lett. {\bf 109},
  026803 (2012).


\bibitem{heiblum}I. Gurman, R. Sabo, M. Heiblum, V. Umansky, and
  D. Mahalu, Nature Communications {\b 3}, 1289 (2012).

\bibitem{us} L. Arrachea and E. Fradkin, Phys. Rev. B {\bf 84}, 235436
  (2011).

\bibitem{capcoup} B. S. Kim, W. Zhou, Y. D. Yash, C. Zhou,
  N. I\c{s}ik, M. Grayson, Rev. Sci. Instruments {\bf 83}, 024703
  (2012).

\bibitem{mcclure} D.T. McClure et al., Phys. Rev. Lett. {\bf 103}, 206806 (2009).

\bibitem{cham-wen}C. de C. Chamon, D. E. Freed, and X. G. Wen,
  Phys. Rev. B {\bf 51}, 2363 (1995).

\bibitem{cham-frad}C. de C. Chamon and E. Fradkin, Phys. Rev. B {\bf
    56}, 2012 (1997).

\bibitem{cappelli} A. Cappelli, M. Huerta, and G. Zemba, Nucl. Phys. B
  \textbf{636}, 568 (2002).

\bibitem{grosfeld} Eytan Grosfeld and Sourin Das, Phys. Rev. Lett.
  \textbf{102}, 106403 (2009).

\bibitem{kovrizhin} D. L. Kovrizhin and J. T. Chalker,
  Phys. Rev. Lett.  \textbf{109}, 106403 (2012).

\bibitem{kane-fis-97} C. L. Kane and M. Fisher, Phys. Rev. {\bf 55},
  15832 (1997).

\bibitem{stern}G. Viola, S. Das, E. Grosfeld, and A. Stern,
  Phys. Rev. Lett. {\bf 109}, 146801 (2012).

\bibitem{Stone}M. Stone, {\em Bosonization}, World Scientific, (1994).

\bibitem{Haldane} F. D. M. Haldane, J. Phys. C: Solid State Phys.,
  {\bf 14}, 2585 (1981).

\bibitem{Karevski-Platini} D. Karevski and T. Platini,
  Phys. Rev. Lett. {\bf 102}, 207207 (2009).

\bibitem{Prosen-Zuncovic} T. Prosen and B. Zuncovic, New J. of
  Phys. {\bf 12}, 025016 (2010).

\bibitem{pathria} R. K. Pathria, {\it Statistical Mechanics}
  (Butterworth-Heinemann Eds., 1996).

\bibitem{eng-an}H. L. Engquist and P.W. Anderson, Phys. Rev. B {\bf
    24}, 1151 (1981).

\bibitem{pendry} J. B. Pendry, J. Phys. A: Math. Gen. {\bf 16}, 2161
  (1983).

\bibitem{lang} See H. Haug and A. P. Jauho, {\it Quantum Kinetics in
    Transport and Optics of Semiconductors} (Springer Series in
  Solid-State Sciences, Vol. 123, Springer-Verlag, Berlin Heidelberg,
  1996).

\bibitem{GR} I. S. Gradshteyn, I. M. Ryzhik, Table of integrals,
  series and products, Academic Press (New York), (1994).


\end{thebibliography}
\end{document}